About explosive delayed desorption from methane-doped argon matrices

M.A. Bludov, I.V. Khyzhniy, S.A. Uyutnov, E.V. Savchenko

[1]Institute for Low Temperature Physics & Engineering, NASU, Kharkiv 61103, Ukraine

e-mail: savchenko@ilt.kharkov.ua

**Abstract**

The features of delayed desorption from $CH_4$-doped Ar matrices irradiated with an electron beam of subthreshold energy were studied. Radiolysis products were detected by emission spectroscopy. The total desorption yield was monitored by recording the pressure in the experimental chamber. Based on the analysis of the concentration dependence of delayed desorption bursts and their structure, an assumption was made about the formation of $CH_4$ clusters in Ar matrices. At a high dopant concentration of 10%, up to three consecutive bursts were recorded. Delayed desorption from a sample doped with 0.1% $CH_4$ was registered for the first time. A correlation has been found between the burst of particles and the flash of luminescence of H atoms. This seemingly contradictory observation of the correlation of the H atom content with the particle explosion due to their recombination was explained by analyzing the energy transfer and capture processes and the features of the energy structure of $H_2$ molecule. A linear dependence of the total particle yield on the $CH_4$ concentration in the range of 1-10% was found. The dynamics of intensity changes in the sequence of main flashes and their delay time were discussed.

**Introduction**

The ongoing interest to the phenomenon of delayed explosive desorption is connected with its possible role in the distribution of matter between gas and solid phase in dense clouds, viz. existence of complex organic molecules (COMs) in the gas phase of cold molecular clouds at temperatures down to 10 K. A number of studies, theoretical modeling and laboratory simulations have been conducted ([1-12] and references therein), and it has been concluded that explosive desorption is the most effective mechanism that can explain the unexpectedly high abundance of some COMs in the gas phase. Delayed explosive desorption is a non-thermal process based on the conversion of energy released in chemical reactions between radiation-induce reagents (radical-radical recombination) into the kinetic energy of the reaction products.

This phenomenon has been observed in various ices of astrophysical interest, in particular in methane [11-14] and methane-containing ices [10, 15-20]. Different kind of ionizing radiation were used – ions in [11, 12], electrons in [10, 13, 14, 16-20] and VUV photons in [15]. In this study mixture of $CH_4$, $CO$, $NH_3$ and $H_2O$ was photolyzed at 10 K. The explosive ejection of material from the sample was triggered by heating. The main explosive event with a loss of 60% of the sample occurred at approximately 27 K with a pressure rise from $10^{-6}$ mb to almost $10^{-4}$ mb, and then several weak pressure surges were observed against the background of a thermal pressure increase. In the experiments [11] with deuterated methane exposed to 1.5 MeV $He^+$ and $H^+$ ions the delayed desorption of $D_2$ was observed at 20 K. Irradiation of methane ices by 9.0 MeV α particles and 7.3 MeV protons at 10 K resulted in the delayed desorption of molecular hydrogen $H_2$ [12] with a loss of about 90 % of the sample. Note, that the recombination of H atoms was considered to be the most efficient mechanism for stimulating desorption [21, 22]. In the study [12], ice with a thickness of 1050±100 nm from a mixture of $CH_4$ and CO was irradiated with an electron beam with an energy of 5 keV and current density of 30 nA $cm^{-2}$. The authors observed the pressure surge (from about $1\ 10^{-10}$ to $3\ 10^{-10}$ torr) after 11 min delay. This event correlated with a decrease in the content of $CH_3$ and HCO radicals and an increase in the content of $C_2H_6$, as well as $H_2CO$ and $CH_3CHO$ species, confirming the efficiency of radical-radical recombination in stimulating delayed desorption. However, the reaction scheme considered did not include recombination of H atoms. The model of delayed desorption based on the recombination of radiolysis products – $CH_3$ and H atoms, was first proposed by Carpenter [23], considering the problem of cryogenic methane moderators of neutron. During long-term irradiation, periodic rises of temperature (by several tens of degrees Kelvin) with hydrogen bursts were detected at intervals of about 24 hours at T=14 K. The destructive one, which resulted in damage to the moderator vessel, occurred after 334 hours at 9 K. Such periodic bursts of temperature represent the case of thermo-concentration self-oscillations [24]. The spontaneous release of energy, which eventually leads to the destruction of the moderator vessel, attracted considerable attention and stimulated further research [16, 18-20, 25–31]. According to Carpenter's model [23] all recombination processes in solid methane moderators proceed in one stage and are controlled by the same activation energy for radiolysis products diffusion 155 K. However, commissioning tests of the ISIS TS2 solid methane moderator, in which a „burp"-like effect was observed [28], and the study of relaxation phenomena in solid methane pre-irradiated with an electron beam [29, 30] led to the conclusion that the radiolysis product recombination process happens in two stages, at different temperatures, and is therefore controlled by two different activation energies $\Delta E_1/k_B = 108.6$ K and $\Delta E_2/k_B$ = 235 K. Modifying Carpenter's approach in this way Kirichek and coauthors received a satisfactory description of the results of the commissioning tests of the ISIS TS2 solid methane moderator. The isochoric pressure derivative equation added to the model [28], allowed to obtain the pressure change in the moderator vessel as a function of time [31].

Strong explosive delayed ejection of particles was observed in solid methane exposed to an electron beam of the subthreshold energy [15, 32] to exclude defect formation and desorption by the impact mechanism. The effect was observed at liquid helium temperature upon reaching a critical irradiation dose of 100 eV per methane molecule and was accompanied by the sample temperature rise and a flash of the luminescence. In contrast to scenario observed under other kind of ionizing radiation – neutrons, ions, superthreshold energy electrons, the burst of particles stimulated by subthreshold energy electrons was preceded by increasing amplitude oscillations of particles desorption. The period of these oscillations depended on the irradiation intensity, it decreased with increasing beam current density [15]. The model proposed in [16] qualitatively describes the occurrence of two types of self-oscillations during electronic excitation of solid $CH_4$, which represent a periodic change in the temperature and concentration of the radiolysis products – H atoms and $CH_3$ radicals, with the irradiation time. The found patterns of self-oscillations was shown to determine the temporal dynamics of the delayed explosive particle desorption upon irradiation of solid $CH_4$ with the subthreshold electrons.

Despite the similarities, there are differences in the manifestations of delayed desorption from irradiated methane-containing ices, which requires further research. To get more insight into the phenomenon the matrix isolation method was introduced in combination with a set of emission spectroscopy methods [18-20, 33]. These studies were carried out in Ar matrix due to effective excitation of the $CH_4$ molecule by free and self-trapped Ar excitons (12.06 and 9.8 eV, respectively [34]) which fall into the range of the dopant absorption band [35]. Moreover, because a small penetration depth of slow electrons (about 100 nm for 1.5 keV electrons [36]) only thin subsurface layer is excited by electrons whereas more than 99% of the sample is excited by excitons and photons (emission of the self-trapped excitons) and the case should be considered as „intrinsic photolysis" [33]. It should be noted that in contrast to pure methane and mixtures of methane with other molecular gases of astrophysical interest there are only a few studies performed in matrices of rare gases– [37] in Kr and [38,39] in Ne matrices. But the delayed desorption was not a topic in these studies. The use of the method of non-stationary luminescence [40] in combination with cathodoluminescence allowed us to prove that the main „actors" in the phenomenon are both particles – H atoms and $CH_3$ radicals. [18-20, 33]. The CH$_3$ radicals were traced via CH emission [33]. In the presented research, we focused on investigating the concentration dependence of delayed desorption from argon matrices doped with methane. Particular attention was paid to maintaining the identical procedures for growing and irradiating samples. It has been proven that the delayed desorption is of bulk nature. This type of desorption was first observed in an Ar matrix with such a low dopant content as 0.1% CH$_4$. A correlation between the particle burst and the luminescence flash of H atoms was discovered and explained considering the processes of energy transfer and capture and the features of the energy structure of H$_2$. At a high dopant concentration of 10%, a series of pressure bursts (up to 3) were detected, which were repeated over long periods of time until the sample was completely sublimated. Each burst was accompanied by a limited number of short-period oscillations A linear dependence of the total particle yield on the CH$_4$ concentration in the range of 1-10% was found. The dynamics of main bursts change in their intensity and the delay time was discussed. It has been suggested that methane clusters are formed in the matrices.

**Experimental**

The experimental procedures have been described in detail elsewhere [18]. Here we present a brief account of the procedures. Particular attention was paid to maintaining the identical procedures for growing and irradiating samples. Ar gas (99.998%) and CH$_4$ gas (99.97%) were used without further purification. Mixture of Ar and CH$_4$ of desired concentration was performed in the gas-handling system which was heated and degassed before each experiment. In these experiments, we performed measurements in the range of CH$_4$ concentrations from 0.1 to 10%. CH$_4$-doped solid Ar films were grown under identical conditions by depositing a certain amount of premixed gas of room temperature onto a liquid helium (LHe)-cooled oxygen-free Cu substrate mounted in a high-vacuum chamber with a base pressure of <$10^{-7}$ torr. The high-quality transparent films had a thickness of 50±10 μm. The irradiation was performed in dc mode with subthreshold energy electrons ($E_e$<$E_{threshold}$ = 1.7 keV) to avoid the knock-on sputtering. In these experiments, the electron beam energy $E_e$ was set to 1.5 keV with the current density of 2.5 mA cm$^{-2}$. The beam covered the icy film with an area of 1 cm$^2$. When the beam was turned on, the pressure slightly increased to 6·$10^{-7}$ torr due to electronically induced desorption of the matrix, which was confirmed by the emission of Ar atoms and unrelaxed molecules. The presence of radiolysis products and impurities was controlled by emission spectroscopy. The spectra were recorded in a wide range – from 50 to 700 nm. The sample temperature was monitored during the entire experiment with a Si sensor mounted on the substrate. The open surface of the samples enabled us to measure cathodoluminescence (CL) spectra not only in visible but also in the vacuum ultraviolet (VUV) range as well as to monitor the particle yield. The total yield of particle desorption under beam was monitored with an ionization detector (a Bayard-Alpert gauge) throughout the entire experiment. Note, that all measurements were performed in the dynamic pumping mode. Pumping out was carried out by LHe cryogenic pump and magnetic discharge pump.

**Results and discussion**

Methane molecule has no intrinsic luminescence spectrum because its excited electronic states are prone to dissociation. Main channels of CH$_4$ dissociation upon photon flux with the threshold wavelengths [41] in the solid Ar emission range are:

$$CH_4 + h\nu \rightarrow CH_3 + H \qquad (1)$$
$$CH_4 + h\nu \rightarrow CH_2 + H_2 \qquad (2)$$
$$CH_4 + h\nu \rightarrow CH_2 + 2H \qquad (3)$$
$$CH_4 + h\nu \rightarrow CH + H + H_2 \qquad (4)$$
$$CH_4 + h\nu \rightarrow C + 2H_2 \qquad (5)$$

CL spectra of solid Ar doped with CH$_4$ contain bands of matrix – emission of the self-trapped excitons Ar$_2^*$ (corresponding to the transition $^{1,3}\Sigma_u^+ \rightarrow {}^1\Sigma_g^+$ [34]), which was recorded in I-st and II-nd order, weak features which belong to Ar atoms and unrelaxed Ar molecules desorbing in excited states. An example of a CL spectrum is shown in Fig. 1 for Ar matrix doped with 0.1 % CH$_4$. The radiolysis products – H atoms – in the sample volume were monitored by the excimer emission band near 166 nm, which corresponds to the emission of the Ar$_2$H* center during the transition to the repulsive part of the ground state potential curve [42]. The desorption of excited H* atoms, which we detected in more concentrated mixtures along the second-order Ly-α line [43], was not observed in the Ar matrix doped with 0.1% CH$_4$ because of the low concentration of the dopant and, consequently, of radiolysis products. As can be seen from the fragmentation reactions (1-5), H atoms are formed in reactions (1), (3) and (4) with the branching ratios BR=0.5: 0.2 and 0.1 respectively for a photon energy of 9.8 eV (self-trapped excitons) and BR=0.25; 0.5 and 0.2 for a photon energy of 12.1 eV (free excitons) according to [44], in which an analysis of energy-dependent branching ratios was presented. It is worth noting that the H atoms formed in these reactions have excess kinetic energy, for example, in channel (1) the experimental average kinetic energy of the lighter H fragment was found to be 3.1 eV [44]. In most other channels this energy exceeds 1 eV and such „hot" H atoms can diffuse over fairly large distances, facilitating atom-atom recombination after thermalization. The mobility of H atoms in an Ar matrix was considered in [45, 46]. Excitation of H atoms by energy transfer leads to the formation of $Ar_2H^*$ excimers, and their radiative transition to the repulsive part of the ground state potential curve again leads to the formation of „hot" H atoms and the resumption of the diffusion process. Detection of molecular hydrogen $H_2$ is hampered because the wide emission continuum of $H_2$, associated with transitions from the bound triplet state $a^3\Sigma_g^+$ to the repulsive lower state $b^3\Sigma_u^+$, overlaps with the $Ar_2^*$-band recorded in the second order. According to [47], the maximum of the spectral distribution of the radiative transition probability from the level v'=0 of the term $a^3\Sigma_g^+$ is about 260 nm. Other radiolysis products – CH radicals, were registered by the emission bands at 387 nm (the $B^2\Sigma^- \to X^2\prod$ transition) and 432 nm (the $A^2\Delta \to X^2\prod$ transition). As was established in [33], CH radicals are „markers" of the methyl radical $CH_3$. C atoms were recorded by the emission line at 295 nm (the $^5S^0 \to {}^3P$ transition) and secondary radiolysis product $C_2$ molecules manifested themselves by emission of Swan bands (channels of $C_2$ formation will be discussed in a further article).

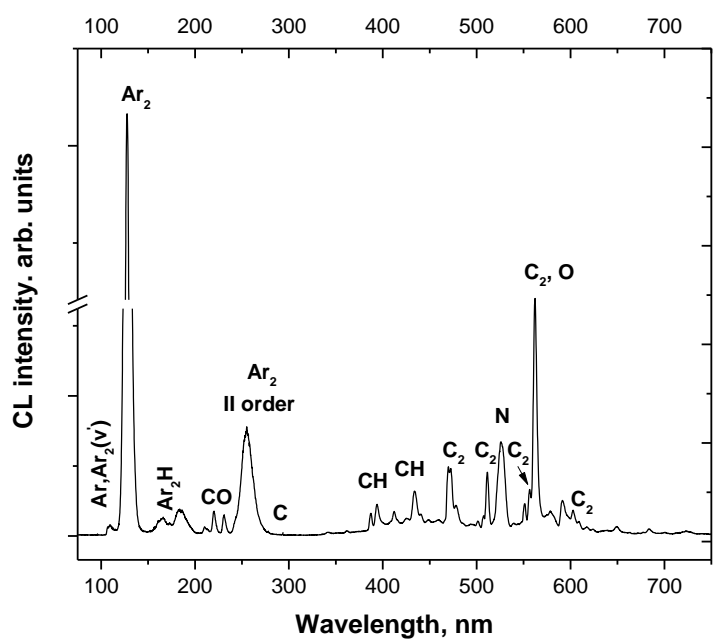

**Fig**. 1. CL spectrum of Ar matrix doped with 0.1 % $CH_4$.

We detected the following impurities: CO (by Cameron bands), N and O atoms (by the $^2D \to {}^4S$ and $^1S \to {}^3P$ transitions correspondingly). Time-correlated measurements of hydrogen atom emission (band at 166 nm) and particle yields were performed during long-term exposure to the electron beam. After approximately one hour of exposure, a luminescence flash and a particle burst were simultaneously recorded as shown in Fig. 2.

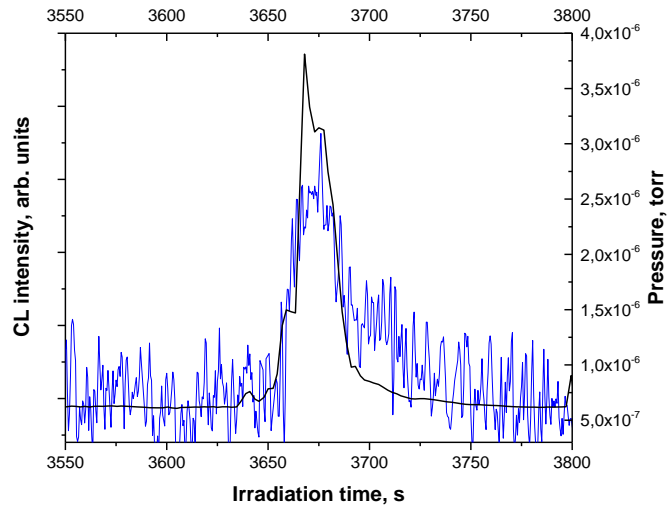

**Fig. 2.** Dose dependence of H atom optical emission and total yield of particles from Ar matrix doped with 0.1 % $CH_4$.

As one can see the delayed desorption correlates with the content of H atoms in the matrix. This seemingly mysterious behavior is explained by the features of the $H_2$ potential energy curves and the mechanism of energy transfer to the $H_2$ molecule. As it was shown in [33] free excitons (12.1 eV) of the matrix can populate low excited electronic states of $H_2$ molecule: the singlet state $B^1\Sigma_u^+$ and the triplet state $a^3\Sigma_g^+$. Transitions from the bound triplet state $a^3\Sigma_g^+$ to the repulsion curve of the lower state $b^3\Sigma_u^+$ result in the dissociation of the hydrogen molecule and the appearance of two „hot" H atoms, which, after thermalization, form centers responsible for the emission of the H band. Thus, the processes of recombination of H atoms to form $H_2$ molecules with the energy release compete with the dissociation of $H_2$ under the action of matrix excitons, and the behavior of the band at 166 nm is due to the contribution of both processes. The effective energy transfer by free excitons to the dopant and radiolysis products is evidenced by the strong quenching of the emission band of self-trapped excitons. Note that the diffusion length $L_{dif}$ of thermalized free excitons in solid Ar ($L_{dif}$~100 nm [48, 49]) exceeds the distance between the dopants even at C=0.1%.

Let us discuss features of the concentration dependence of delayed desorption. The delay time τ of the first burst relative to the start of irradiation is almost independent of the methane concentration, and approaches an hour: $\tau_{0.1}$= 3670±100 s, $\tau_1$=3550 ±100 s, $\tau_5$= 3530±100 s and $\tau_{10}=3300\pm100$ s (the lower index in $\tau$ corresponds to the concentration of $CH_4$). The sequence of bursts is shown in Fig. 3.

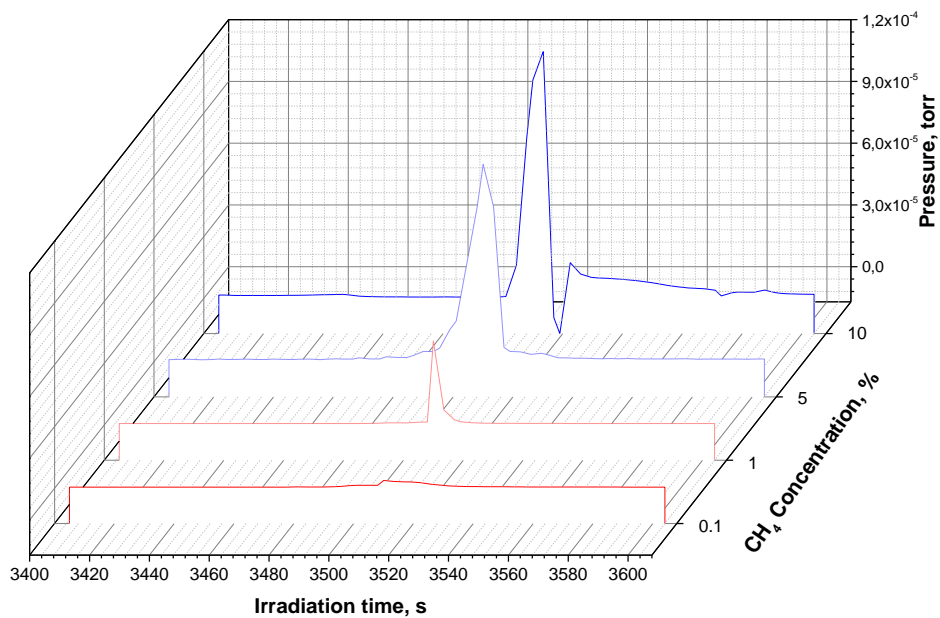

**Fig. 3.** The first bursts of delayed desorption under the action of an electron beam irradiating Ar matrices doped with $CH_4$ with a concentration of C=0.1; 1; 5 and 10%. For ease of comparison, the burst curves are reduced to the same delay time.

In fact, the delay time for the lowest concentration $\tau_{0.1}$ is slightly longer than the average delay time, and the delay time for the highest concentration $\tau_{10}$ is slightly shorter than this time. The details of the burst contours observed on samples with different concentrations are presented in Figs. 5, 6, 8-10. The independence of the delay time $\tau$ from the concentration of $CH_4$ suggests that the methane molecules form clusters in the Ar matrices. Methane molecules occupy substitutional cavities in Ar at low concentration of $CH_4$. The substitutional site diameter in Ar matrix is 3.755 Å [50], as compared to 3.988 Å [51], the diameter of the $CH_4$, molecule. Thus, the molecule is squeezed into smaller cavity in Ar especially at low temperatures. A study [52] of the stretching vibration of $CH_4$, which is very sensitive to the environment, showed that the distribution of molecules in rare gas matrices is not random and that they are distributed in an ordered manner to minimize the strain energy. In the Ar-$CH_4$ system, the $v_3$ vibration band of $CH_4$ clearly demonstrated that at a $CH_4$ concentration of only 3%, the solid changed the inclusion structure. The number of certain types of clusters – dimers and trimers, in a crystal formed of two types of atoms/molecules has been calculated in different lattices, taking into consideration only interactions between nearest neighboring atoms [53]. The data for the FCC lattice at different concentration, reconstructed from Table III of [53], are presented in Fig. 4 considering both dimers and trimers. At a $CH_4$ concentration of 0.1%, only about 1% of atoms can be found in dimers, but at a $CH_4$ concentration of 5%, 32% of molecules can occupy adjacent positions in dimers and trimers. This works towards the formation (in reaction 1) of $CH_3$ in close proximity, facilitating their recombination with the release of energy. A concentration of 10% is the limit at which the mixture splits into separate FCC phases of argon and methane [54].

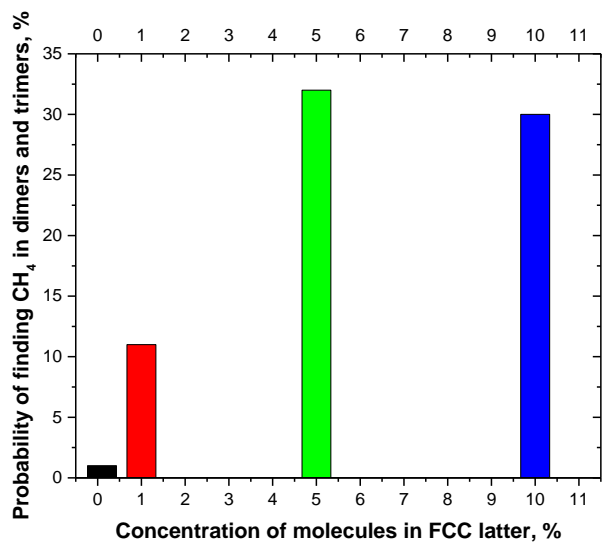

**Fig. 4** Probability of finding $CH_4$ in dimers and trimers at different concentration.

The dynamics of particle bursts under the action of an electron beam is shown using the example of a sample doped with 10% $CH_4$, in which 3 bursts of particle desorption with decreasing intensity were observed. (Fig. 5). Note that the delay time to the next burst decreases. So, if the delay time $\tau^1_{10}$ to the first burst was 3300±100 s, the delay time $\tau^2_{10}$ to the second burst was $\tau^2_{10}$=600±50 s, and to the third one $\tau^3_{10}$=300±50 s.

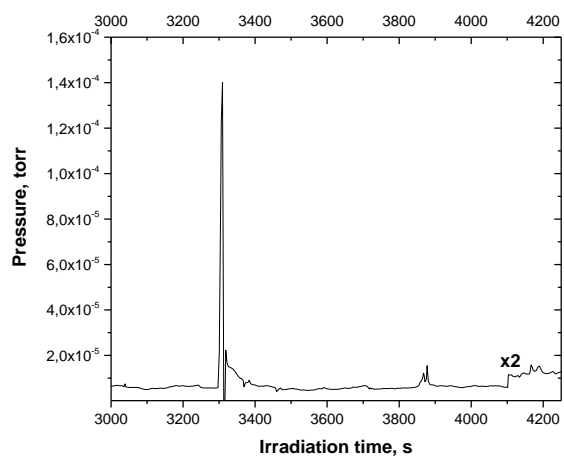

**Fig. 5.** Delayed explosive desorption of Ar matrix doped with 10% $CH_4$ traced by pressure.

During the first burst, we observed an increase in temperature shown in Fig. 6, indicating exothermicity of the process. It should be noted that since the temperature sensor was mounted on the back of a substrate efficiently cooled by LHe, it was not possible to determine the actual temperature of the Ar film.

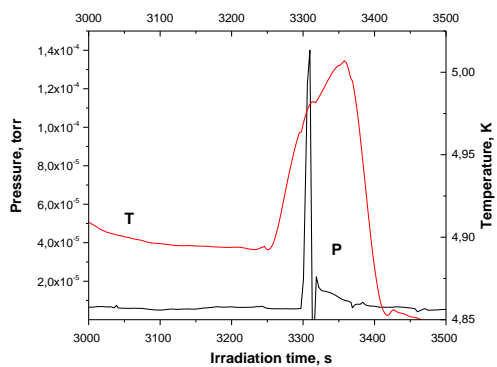

**Fig. 6.** First pressure burst of delayed desorption from Ar matrix doped with 10% $CH_4$ and temperature of the sample recorded in correlation manner by Si sensor.

The beginning of sublimation of the sample allows one to estimate the lower limit of the heating pulse — 30 K. If we compare the conditions of a series of particle bursts, it becomes clear that the conditions in each series are different. During the first burst, a lot of the sample mass evaporates (about 90% by our estimation). Because the samples have an open surface and the experiment is conducted in dynamic pumping mode, desorbed particles practically do not recondense on the sample after the explosion. Therefore, much less material is available for the second explosion, and the surge is reduced in intensity. On the other hand, due to the decrease in the number of molecules in the sample after the explosion, the delivered radiation dose per molecule increases, resulting in a decrease in the desorption delay time. This process is repeated, and the corresponding changes occur after the second pressure surge. With decreasing concentration, i.e. the number of active methane centers, the mass of the sample that sublimates during the first burst decreases according to a linear law, as shown in Fig. 6. From this series, only the sample with the lowest concentration of $CH_4$ – 0.1%, falls out, during irradiation of which the pressure increases only from $6.4 \times 10^{-7}$ to $3.7 \times 10^{-6}$ torr.

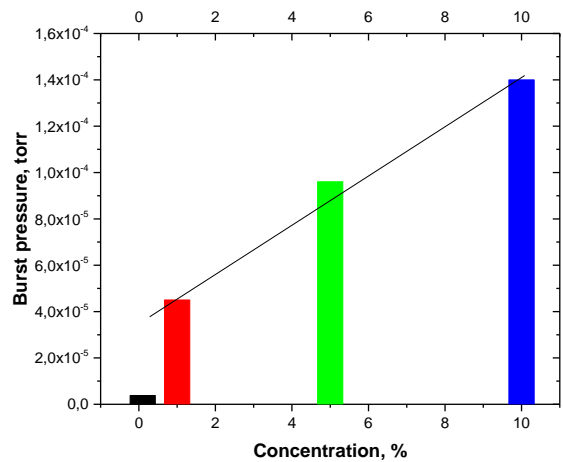

**Fig. 7**. Concentration dependence of delayed desorption of Ar matrices doped with methane.

The heat of sublimation $\Delta H$ for Ar is $\Delta H= 7.79$ kJ mol$^{-1}$ according [55]. The estimated number N of particles in the matrix is N ~ $10^{20}$ and the required energy $E_s$ to sublimate the entire sample is about 1 J. Considering the energy release $\Delta E_H$ upon recombination of H atoms $\Delta E_H = 218$ kJ mol$^{-1}$ and $\Delta E_{CH3}$ upon recombination of $CH_3$ radical $\Delta E_{CH3} = 368$ kJ mol$^{-1}$ we can roughly estimate of the critical number n of the radical concentration needed to sublimate the entire sample. It turned out that n ~ $10^{-2}$ is required to evaporate the entire sample, and less, respectively, to evaporate some part of the sample. The concentration dependence of the delayed desorption of Ar matrices, shown in Fig. 7, reflects this trend.

The different dynamics of pressure surges in our experiments and under operating conditions of methane moderators is most likely due to the difference in experimental conditions. As it was mentioned, our samples which have an open surface lost a significant part of material in bursts of the delayed desorption. Moreover, some small part of the material is converted to carbon compounds found on the substrate after numerous experiments. In cryogenic moderators, methane is contained in a closed vessel and after the temperature and pressure burst the system returns to its initial state practically without any essential loss of material if did not take into consideration the production of a polymer mixture in vessel via radiolysis reactions discussed in [31].

A feature of the delayed particle desorption in methane-doped Ar matrices is a limited number of short-period oscillations that accompany the long period bursts. Examples of such bursts are shown in Fig. 5 and 6 for the matrix containing 10% of $CH_4$. In the Ar matrix doped with 5% $CH_4$ we also observed sharp rise in the total yield of particles after about an hour exposure to an electron beam, which was preceded by short-period oscillations as shown in

Fig. 8. It is worth noting that the period of rapid oscillations is almost the same as that in pure methane.

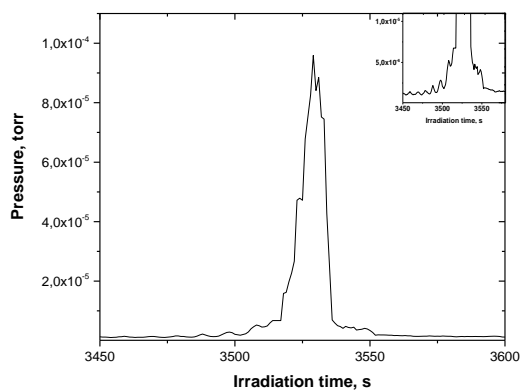

**Fig. 8.** The first burst of particles in delayed desorption from the Ar matrix doped with 5% $CH_4$. Preceded oscillations are shown in the inset.

In lightly doped matrices (1% $CH_4$) a limited number of short-period oscillations was detected as shown in Fig. 9.

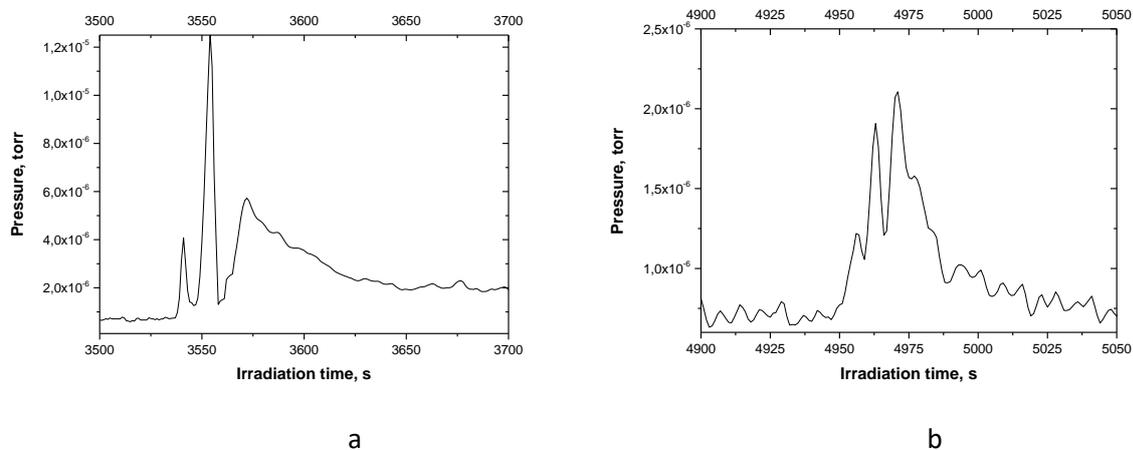

a                                          b

**Fig. 9** Structure of the first burst (a) and second burst (b) from Ar doped with 1 % $CH_4$ observed in [18] (with kind permission of FNT/LTP).

The structure observed during the flare at the lowest $CH_4$ concentration of 0.1% is blurred, as can be seen in Fig. 10, but still noticeable.

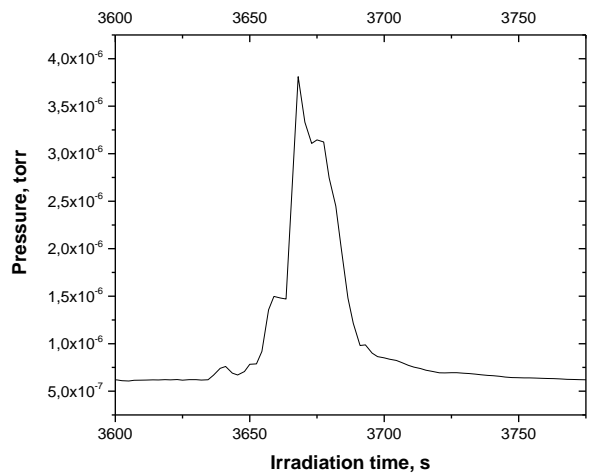

**Fig. 10.** Structure of the particle burst from Ar matrix doped with 0.1% of $CH_4$.

According to the theoretical calculations performed for pure methane in [56], the number of short-period oscillations is defined by radiation-stimulated degradation of methane molecules, that is involvement of methane molecules in various channels of radiolysis. Considering that our data indicate the formation of methane clusters in matrices, we can expect a similar effect of reducing the number of short-period oscillations in matrix systems which can be considered as thermo-concentration instability. However further efforts are needed to deepen our understanding of the phenomenon.

**Summary**

The features of delayed desorption from argon matrices doped with 0.1 – 10% methane and exposed to an electron beam were studied using the emission spectroscopy methods. Following radiolysis products were detected – H atoms, CH radicals, C atoms and $C_2$ molecules. Desorption was monitored by recording the pressure in the experimental chamber. Delayed desorption from a sample doped with 0.1% $CH_4$ was registered for the first time. A luminescence flash of H atoms was observed after nearly an hour of irradiation which correlated with explosive pulse of particle ejection. Consideration of the energy transfer and potential energy curves of the $H_2$ molecule made it possible to explain the seemingly contradictory observation of a burst of optical emission of H atoms as an indicator of their recombination. Based on the analysis of the concentration dependence of delayed desorption bursts and their structure, an assumption was made about the formation of methane clusters in Ar matrices. The study of the dose dependences of desorption over a long time until the complete evaporation of the sample allowed us to find a series of pressure bursts (up to 3). A linear dependence of the intensity of delayed desorption on the concentration of $CH_4$ molecules in the concentration range of 1 - 10% $CH_4$ in the Ar matrix was established. The appearance of a limited number of short-period oscillations along with explosive bursts of delayed desorption, which occur after long periods of time, was demonstrated. The features of the dynamics of a series of particle desorption bursts were discussed – a decrease in the intensity of each subsequent burst and the delay time, as well as the appearance of a limited number of short-period oscillations. We hope that the presented results will stimulate further studies of delayed desorption from systems containing reactive species.

**Acknowledgements**

The authors cordially thank colleagues Vladimir Sugakov, Oleg Kirichek, Giovanni Strazzulla, Anatoli Popov and Hermann Rothard for stimulating discussions.